\documentclass[submission,copyright,creativecommons]{eptcs}
\providecommand{\event}{Postproceedings of ITRS'14}
\usepackage{amssymb}
\usepackage{amsthm}
\usepackage{epsfig}
\usepackage{amsmath}
\newcommand{\ra}{\rightarrow}

\newcommand{\Lra}{\Leftrightarrow}
\newcommand{\Ra}{\Rightarrow}

\newcommand*{\doi}[1]{\href{http://dx.doi.org/\detokenize{#1}}{doi: \detokenize{#1}}}

\begin{document}

\pagestyle{headings}

\title{A Finite Model Property for Intersection Types}

\def\titlerunning{A Finite Model Property for Intersection Types}

\author{Rick Statman
\institute{Carnegie Mellon University\\
Department of Mathematical Sciences\\
Pittsburgh, PA  15213}
\email{statman@cs.cmu.edu}}

\def\authorrunning{Rick Statman}
\maketitle

\begin{abstract} We show that the relational theory of intersection types known as BCD has the finite model property; that is, BCD is complete for its finite models.  Our proof uses rewriting techniques which have as an immediate by-product the polynomial time decidability of the preorder $\subseteq $ (although this also follows from the so called beta soundness of BCD).
\end{abstract}

\section{Barendregt, Coppo, and Dezani}

BCD is the relational theory of intersection types presented by Henk Barendregt, Mario Coppo, and Mariangiola Dezani in \cite{2}.  Here we consider the theory, without top element, as about a preorder $\subseteq$,

\vspace{.15in}

\noindent $a \subseteq a$

\noindent $a \subseteq b\   \& \ \subseteq c \Ra a \subseteq c$

\noindent $a \wedge b \subseteq a$

\noindent $a \wedge b \subseteq b$

\noindent $c \subseteq a\ \&\ c \subseteq b \Ra c \subseteq a \wedge b,$

\vspace{.15in}

\noindent and a contravariant-covariant operation $\ra$,

\vspace{.15in}

\noindent $c \subseteq a \ \&\ b \subseteq d \Ra a \ra b \subseteq c \ra d$

\vspace{.15in}

\noindent satisfying the weak distributive law

\vspace{.15in}

\noindent $(c\ra a) \wedge (c\ra b)\ \subseteq  c \ra (a \wedge b).$

\vspace{.15in}

Of course it is well known that if the points of such a preorder are partitioned by the congruence $\sim$ defined by

\vspace{.15in}

\noindent $ a \sim b \Lra a \subseteq b  \ \& \ b \subseteq  a $

\vspace{.15in}

\noindent we obtain a semilattice with $\wedge$, that is,

\vspace{.15in}

\noindent $\begin{array}{rcl}  a \wedge (b \wedge c) & \sim & (a \wedge b) \wedge c \\
a \wedge b & \sim &  b \wedge a \\
a & \sim & a \wedge a
\end{array}$

\vspace{.15in}

\noindent where the quotient partial order can be recovered

\vspace{.15in}

\noindent $a \subseteq b\ \Lra \ a \sim a \wedge b.$

\vspace{.15in}

\noindent In addition, the quotient satisfies the distributive law

\vspace{.15in}

\noindent $c \ra (a \wedge b) \sim (c \ra a) \wedge (c \ra b)$

\vspace{.15in}

\noindent and an absorption law

\vspace{.15in}

\noindent $a \ra b \sim (a \ra b) \wedge ((a \wedge c) \ra b).$

\vspace{.15in}

Now if a semilattice is given and $a \subseteq b$ is defined by

$$
a \subseteq b \Lra a = a \wedge b
$$

\noindent then $\subseteq$ is a preorder with a meet operation.
In addition, if the distributive law and the absorption law are satisfied then the $\ra$ operation enjoys the contravariant-covariant property.  There is also a derived absorption law

\vspace{.15in}

\noindent $c \ra (a \wedge b) = (c\ra a) \wedge (c \ra (a\wedge b))$

\vspace{.15in}

\noindent which proves useful.  In this way we have an equational presentation of BCD.

\section{Expressions and their rewriting}

We define the notion of an expression as follows.  @,$p,q,r,\ldots$ are atomic expressions.  If $A$ and $B$ are expressions then so are $(A \ra B)$ and $(A \wedge B)$.  Even though we write infix notation we say that these expressions begin with $\ra$ and $\wedge$ respectively.  The notions of positive, negative, and strictly positive are defined recursively by

\vspace{.15in}

\noindent $A$ is positive and strictly positive in $A$.

\noindent If $C$ is positive in $B$ then $C$ is positive in $A\ra B$ and negative in $B \ra A$.

\noindent If $C$ is strictly positive in $B$ then $C$ is strictly positive in $A\ra B$.

\noindent If $C$ is positive in $A$ or $B$ then $C$ is positive in $A \wedge B$.

\noindent If $C$ is strictly positive in $A$ or $B$ then $C$ is strictly positive in $A \wedge B$.

\noindent If $C$ is negative in $B$ then $C$ is negative in $A \ra B$ and positive in $B \ra A$.

\noindent If $C$ is negative in $A$ or $B$ then $C$ is negative in $A \wedge B$.

\vspace{.15in}

A single occurrence of $B$ as a subexpression of $A$ will be indicated $A[B]$.  An  expression can be thought of as a rooted oriented binary tree with atoms at its leaves and either $\ra$ or $\wedge$ at each internal vertex.  For each subexpression $B$ of $A$ there is a unique path from the root of $A$ to the root of $B$.  The ebb of $B$ in $A$ is the number of $\ra$ verticies on the path from the root of $A$ to the root of $B$; so $C \ra D$ has ebb $=1 $  in $C \ra D$.  The $\ra$ depth of $A$ is the maximum ebb of a subexpression of $A$.

With an equational presentation we can associate a set of rewrite rules.  The one step rewrite of an expression $A$ by the rule $R$ to the expressiion $B$ is denoted $A$ $R$ $B$.  This is the replacement of exactly one occurrence of the left hand side of the rule as a subexpression of $A$, the redex, by the right hand side.  Sets of rules can be combined by the regular operations $+$ (union) and $*$ (reflexive-transitive closure).  Now fix $n$ to be a natural number or infinity $= o$(mega).  We define rewrites

\vspace{.15in}

\noindent \begin{tabular}{lrcl}
(asso.) & $A \wedge (B \wedge C)$ & asso. &$(A \wedge B) \wedge C$\\
(asso.) & $(A \wedge B) \wedge C$ & asso. & $A \wedge (B \wedge C)$\\
(comm.) & $A \wedge B$ & comm. & $B \wedge A$\\
(idem.) & $A$ & idem. & $A \wedge A$ \\
(absp.) & $A \ra B$ & absp. & $(A \ra B) \wedge ((A \wedge C) \ra B)$\\
(dist.) & $A \ra (B \wedge C)$ & dist. & $(A \ra B) \wedge (A \ra C$\\
(dept.) & $A[B]$ & dept. & $A [@]$ if $B$ lies at ebb $> n$ in $A[B]$
\end{tabular}

\vspace{.15in}

\noindent and we set semi. $=$ asso. $+$ comm., and slat. $=$ semi. $+$
idem.  Let\newline
 redn. $=$ slat. $+$ absp. $+$ dist. $+$ dept.\ .  Of course, when $n =$ infinity dept. is trivial and redo. generates the congruence on expressions induced by BCD.

Given a reduction $A(1)$ redn. $\ldots$ redn. $A(k)$ an occurrence of $\ra$ in $A(k)$ has a unique ancestor in each $A(i)$ except when $A(i)$ absp. $A(i+1)$ and the $A(i+1)$ ancestor of $\ra$ lies in $C$.  Similarly for atoms.

\section{Properties of the rewriting system}
\begin{enumerate}
\item[(1)]  idem. can be restricted to atoms.

\begin{proof} $A \wedge B$ idem. $(A \wedge A) \wedge B$ idem. $(A\wedge A) \wedge (B \wedge B)$ asso.* $ A \wedge ((A \wedge B) \wedge B)$ comm. $A \wedge ((B\wedge A) \wedge B)$ asso.* $(A \wedge B) \wedge (A \wedge B).\  A \ra B$ idem. $A \ra (B \wedge B)$ dist.   $(A \ra B) \wedge (A \ra B)$.  End of proof.
\end{proof}

\item[(2)] comm. can be restricted to atoms and expressions beginning with $\ra$.

\begin{proof} All permutations can be done by adjacent transpositions.
End of proof.  \end{proof}

\item[(3)]  dept. can be restricted to intersections of atoms and $\ra $ of @'s.

\begin{proof} If $B$ lies at ebb $>n$ in $A$ then any longest $\ra$ path in $B$ ends in an intersection of atoms.  Indeed, since it is longest, it is either $C$ or $D$ in a subexpression $C \ra D$ of $B$, where the other of $C$ and $D$ is similar.  If such an intersection is non-trivial or $p$ it can be replaced by @.  Similarly for the other.  Otherwise, we have a subterm @ $\ra$ @ of $B$ which dept. @.  End of proof.
\end{proof}
\end{enumerate}

From here on we assume that the restrictions in (1), (2), and (3) are obeyed in all reductions.

\begin{enumerate}
\item[(4)] Every dist. reduction terminates.

\begin{proof} The ebb of $\wedge$'s decreases.  End of proof.
\end{proof}

\item[(5)]  Every dept. reduction terminates.

\begin{proof} Either length decreases or atoms change to @.  End of proof.
\end{proof}

\item[(6)]  idem. expedition.

\noindent If $A$ slat.$^*\ B$ then there exists $C$ such that

$$ A \ {\rm idem.}^*\ C\ {\rm semi.}^*\ B
$$

\begin{proof} Each idem. redex has a unique ancestor in $A$ to which idem. can be applied.  End of proof.
\end{proof}

\item[(7)]  dept. postponement.

\noindent If $A$ dept.$^*\ B\ R^*\ C$, where $R \in\  $\{slat., dist., absp.\} then there exists $D$ s.t.

$$
A\ R^*\ D\ {\rm dept.}^*\ C.
$$

\begin{proof} A dept. redex is either an intersection of atoms or @ $\ra$ @.  It has either one or two descendants in the result of any $R$ reduction these are also dept. redexes.  End of proof.
\end{proof}

\item[(8)]  dist. has the weak diamond property.

\vspace{.15in}

\item[(9)] dept. has the weak diamond property.

\vspace{.15in}

\item[(10)]  Parallel moves lemma.

\begin{proof} This lemma has the form: if $A\ R^*\ B$ and $A\ S^*\ C$ then, for some $D,\ B\ S^*\ D$ and $C$ redn.$^*D$ for various $R,S \in $ \{slat., absp., dist., dept.\} so there are 16 possible cases.  We denote these cases $R/S$.  There are several exceptional cases; these are 10.9, 10.13, and 10.14.  These cases must be accounted for separately so they fit together in  a strip lemma argument for the Church-Rosser theorem.  We begin with the special case $R$/idem. which is trivial.

\vspace{.15in}

\noindent (10.1-4) slat./S

\vspace{.15in}

\begin{tabular}{rl}
slat./slat.; & by idem. expedition.\\
slat./dist.; & If $A$ idem.$^*\ B$ and $A$ dist.$^*\ C$ then there exist $D,E$\\
& such that $B$ dist.$^*\ D,\ C$ idem.$^*\ E$ and $D$ semi.$^*\ E$.\\
& If $A$ semi.$^*\ B$ and $A$ dist.$^*\ C$ then there exists $D,E$\\
& such that $C$ dist.$^*\ D,\ B$ dist.$^*\ E$ and $D$ semi.$^*\ E.$\\
slat./absp.; & The strong diamond property holds for sets of \\
& non-overlapping redexes.\\
slat./dept.; & by idem. expedition.\\
\end{tabular}

\newpage

\noindent (10.5-7) remaining R/R

\vspace{.15in}

\begin{tabular}{ll}
dist./dist.; & (4) and (8) give us the strong diamond property.\\
dept./dept.; & (5) and (9) give us the strong diamond property.\\
absp./absp.; & If $A$ absp.* $B$ and $A$ absp.* $C$ then there exists $D,E$\\
& such that $B$ absp.* $D$, $C$ absp.* $E$ and $E$ semi.* $D$
\end{tabular}

\vspace{.15in}

\noindent (10.8-9) remaining R/dist.

\vspace{.15in}

\begin{tabular}{ll}
absp./dist.; & If $A$ absp.$^*\ B$ and $A$ dist.$^*\ C$ then there exists $D,E$\\
& such that $B$ dist.$^*\ D$, $C$ absp.$^*\ E$ and $E$ semi.$^*\ D$\\
dept./dist.; & If $A$ dept.$^*\ B$ and $A$ dist.$^*\ C$ then there exists $D,E$\\
& such that $A$ idem.$^*\  D$ dept.$^*\ E$ and $B$ dept.$^*\ E$. This \\
& is an exceptional case.
\end{tabular}

\vspace{.15in}

\noindent (10.10-11) remaining R/dept.

\vspace{.15in}

\begin{tabular}{ll}
dist./dept.; & This is the same as 10.9 but here it is not exceptional.\\
absp./dept.; & If $A$ absp.* $B$ and $A$ dept.* $C$ then there exist $D,E$ such\\
& that $B$ dept.* $E$, $C$ absp.* $D$, and $D$ idem.* $E$.
\end{tabular}

\vspace{.15in}

\noindent (10.12-13) remaining R/absp.

\vspace{.15in}

\begin{tabular}{ll}
dist./absp.; & This is the same as 10.8 and is not exceptional since\\
& semi. is bidirectional.\\
dept./absp.; & There is one special case which is exceptional. @ $\ra$ @\\
& dept. @ and @ $\ra$ @ absp. (@ $\ra$) $\wedge$ ((@ $\wedge C$) $\ra$ @) so @ idem.\\
& @ $\wedge$ @ and (@ $\ra$) $\wedge$ ((@ $\wedge C$) $\ra$ @) dept.$^*$  @ $\wedge$ @. So, in general \\
 & $A$ dept.$^*\ B$ and $A$ absp.$^*\ C$ then there exist $D$\\
&that $B$ (absp. $+$ idem.)$^*\ D,\ C$ dept.$^*\ D$.
\end{tabular}

\vspace{.15in}

\noindent (10.14-16) remaining R/slat.

\vspace{.15in}

\begin{tabular}{ll}
dist./slat.; & First consider the case $A$ dist. $B$ and $A$ semi.  $C$.  Then\\
& there exists $D$, $E$ such that $B$ dist.* $D$, $C$ dist.* $E$ and \\
& $D$ semi.* $E$.  Now use idem. expedition.  This is an \\
& exceptional case.\\
absp./slat.; & As in 10.3.\\
dept./slat.; & As in 10.4.
\end{tabular}

\vspace{.15in}

\noindent End of proof.
\end{proof}

\noindent We may divide redn. reductions into alternating segments slat.*, absp.*, dist.*, and dept.*.  Such a reduction has the pointedness property if

\vspace{.15in}

\begin{tabular}{ll}
(pointedness) & Every dist.* segment ends in a dist. normal form and\\
&every dept.* segment ends in a dept. normal form
\end{tabular}

\vspace{.15in}

\noindent A reduction is said to be focused if

\vspace{.15in}

\begin{tabular}{ll}
(focus) & The reduction has the pointedness property and every\\
&  segment either ends with a dist. and dept. normal form\\
& or is followed by a dist. segment and then a dept. segment\\
& or vice versa.
\end{tabular}

\item[(11)] Focus lemma.

\vspace{.15in}

If $A$ redn.* $B$ then there is a dist., dept. normal form $C$ of $B$ and a focused reduction from $A$ to $C$.

\begin{proof} Given a reduction from $A$ to $B$ repeatedly apply parallel moves R/dept. to the segments of the reduction where the dept.* is to normal form.  Now repeatedly apply parallel moves R/dist. to the segments of the reduction where the dist.* is to normal form.  In the exceptional case 10.9 we have an expression $X$ reduced on the one hand to dept. normal form $Y$ and on the other hand reduced to dist. normal form $Z$.  Thus the dist. normal form $W$ of $Y$ has $W$ idem.* $Z$.  Now we continue the process with $W$.  In the end all the extra idem.*'s are pushed to the end.  Note that this does not change the status of the final expression although an extra reduction to dist. normal form could be added anyway. End of proof.
\end{proof}

\item[(12)]  Strip lemma.

\vspace{.15in}

\noindent   If $A$ redn.$^*\ B$ by a focused reduction and $A\ R^*\ C$ for $R
\in$  \{slat., dist., dept.\} then there exists $D$ such that $C$ redn.$^*\ D$ and $B\ S^*\ D$ for $S \in$  \{slat., dist., dept.\}.  In addition, if $A$ redn.$^*\ B$ by a focused reduction and $A$ idem.$^*$ absp.$^*\ C$ then there exists $D$ such that $C$ redn.$^*\ D$  and $B$ idem.$^*$  absp.* $D$.

\begin{proof} We have divided redn.$^*$ into alternating segments slat.$^*$, absp.$^*$, dist.$^*$, and dept.$^*$.  The proof is by induction on the number of such segments.  Clearly it suffices to assume that if $R$ is dist. then $C$ is in dist. normal form and similarly for dept.  The basis case is just the parallel moves lemma together with the observation that

\begin{enumerate}
\item[(i)] If the case is 10.9 then the $R$ changes to slat.
\item[(ii)] If the case is 10.13 then the hypothesis considers the exception.\
\item[(iii)]  If the case is 10.14 then $B = D$ since $B$ is dist. normal and the case is not really exceptional.
\end{enumerate}

\noindent For the induction step we suppose that $A\ S^*\ B^{\prime}$ redn. $B$.  We can apply the basis step to $A\ S^*\ B^{\prime}$, and we can apply the induction hypothesis to the reduction $B^{\prime}$ redn.$^*\ B$.  Since the original reduction was pointed these compose to give the result.
\end{proof}

\item[(13)]  Church-Rosser property.

\begin{proof} Let conv. be the congruence generated by redn.  We need to show that if $A$ conv. $B$ then there exists $C$ such that $A$ redn.$^*\ C$ and $B$ redn.$^*\ C$.  The proof is by induction on the length of a conversion from $A$ to $B$.  With the strip and focus lemmas completing the proof is routine.  End of proof.
\end{proof}
\end{enumerate}

\section{The models $F(n)$}

\begin{enumerate}
\item[(14)]  Conservation lemma.

\vspace{.15in}

\noindent If the $\ra$ depth of $A$ is $< n+1$, $A$ redn.* $B$, and $B$ is in dept. normal form then $A$ redo.* $B$.

\begin{proof}  By dept. postponement we may assume that the reduction $A$ to $B$ has all dept. reductions at the end and we have a $C'''$ such that $A$ redo.* $C'''$ dept.* $B$.  Now the dept. redex of $C'''$ contracted next lies in a subexpression $C' \ra C''$ of ebb at least $n+1$ in $C'''$.  Since $A$ has $\ra $ depth $< n+1$  the subexpression $C' \ra C''$ has a unique ancestor which is a subexpression of the $C$ occurring on the right hand side of the absp. reduction rule applied to some redex in the reduction of $A$ to $C'''$.  Now every descendant of this ancestor has ebb at least $n+1$ so the choice of $C$ can be modified to the result of replacing the ancestor  subterm by @, without changing the dept. normal form. End of proof.\end{proof}

Now the conv. congruence has an equational presentation and thus a free model $F(n)$ consisting of congruence classes of expressions.  We adopt the customary notation $F(n)\ \models A = B$ to signify that $A$ and $B$ belong to the same congruence class of $F(n)$.

The stack of 2's function $s(n,m)$ is defined by

\vspace{.15in}

\noindent $\begin{array}{rcl}
s(0,m) & = & m \\
s(n+1,m) & = & 2^{s(n,m)}
\end{array}
$

\vspace{.15in}

\item[(15)] Finiteness lemma.

\vspace{.15in}

\noindent If there are $m$ atoms and $n$ is finite then $F(n)$ has at most $s(n+1,m+n)$ elements.

\begin{proof} It suffices to over estimate the number of dept. normal forms. End of proof.
\end{proof}

\item[(16)] Completeness of the $F(n)$

\vspace{.15in}

If the $\ra $ depths of $A$ and $B$ are both $< n+1$ then $F(n)\ \models A = B$ implies $A$ and $B$ are congruent in BCD.

\begin{proof} Suppose that $F(n)\ \models A = B$.  Then $A$ conv. $B$ so by the Church-Rosser theorem there exists $C$ such that $A$ redn.$^*\ C$ and $B$ redn.$^*\ C$.  By (5) we can assume $C$ is dept. normal.  Thus, by the conservation lemma we have both $A$ redo.$^*\ C$ and $B$ redo.$^*\ C$ so $A$ and $B$ are congruent in BCD. End of proof.
\end{proof}
\end{enumerate}

\section{Polynomial time decidability of $\subseteq$}

First we remark that the beta soundness lemma (\cite{1}) is a simple consequence of the Church-Rosser theorem.

\begin{enumerate}
\item[(17)]  Weak standardization of redo.

\vspace{.15in}

\noindent If $A$ redo.* $B$ then there is a reduction from $A$ to $B$ where no strictly positive redex is contracted after one which is not strictly positive.

\begin{proof} Is straightforward.  End of proof.
\end{proof}

If $A$ has no strictly positive dist. redex then $A$ is an intersection, under some association, of expressions of the form $A(1) \ra (\ldots (A(t) \ra p) \ldots )$ where $p$  is an atom (here we do not distinguish @). We call these expressions the factors of $A$.  We define the set of factors of an expression $E$ more generally by recursion

\vspace{.15in}

\begin{tabular}{lcl}
factors$(p)$ & = & \{ p\}\\
factors$(E'\wedge E'')$ & = & factors $(E')$ \ $\cup$ factors$(E'')$ \\
factors$(E' \ra E'')$ & = & $\{E' \ra E'''\ |\ E'''$ : factors$(E'') \}$
\end{tabular}

\vspace{.15in}

\item[(18)] Complete invariants lemma.

\vspace{.15in}

\noindent If there is a strictly positive reduction from $E'$ to $E''$ then each factor of
$E'$ is a factor of $E''$ and for each factor

$$
E''(1) \ra ( \ldots (E''(t) \ra p )\ldots )
$$

\noindent of $E''$ there exists a factor

$$
E'(1) \ra (\ldots (E'(t) \ra p) \ldots )
$$

\noindent of $E'$ and expressions $D(1) , \ldots , D(t)$ such that  $E''(i)$ slat.* $E'(i)
\wedge D(i)$ for $i = 1, \ldots , t$.

\vspace{.15in}

\noindent {\it Remark.} If we allow each $D(i)$ to be empty then slat. can be replaced  by assoc.

\begin{proof} By inspection of the rewrite rules.  End of proof.
\end{proof}

\item[(19)] Beta soundness (\cite{1}, \cite{4} Lemma 2)

\vspace{.15in}

If both $A$ and $B$ have no strictly positive dist. redexes and $A$ conv. $B$ then for each factor $A(1) \ra (\ldots(A(t) \ra p) \ldots )$ of $A$ there exists a factor $B(1) \ra ( \ldots (B(t) \ra p) \ldots )$ of $B$ and expressions $C(1), \ldots , C(t)$ such that $A(i)$ conv. $B(i) \wedge C(i)$ for $i=1, \ldots , t$.

\begin{proof}  By the Church-Rosser theorem there exists $C$ such that both $A$ and $B$ redo.* $C$. By (4) we may assume that $C$ has no strictly positive dist. redex and by weak standardization there exist $A',B'$ such that

\vspace{.15in}

\noindent $A$ redo.* $A'$ by only strictly positive reductions,

\noindent $B$ redo.* $B'$ by only strictly positive reductions,

\noindent $A'$ redo.* $C$ with no strictly positive reductions, and

\noindent $B'$ redo.* $C$ with no strictly positive reductions.

\vspace{.15in}

\noindent In particular, $A'$ and $B'$ have no strictly positive dist. redexes so their factors are actually subexpressions.  By the complete invariants lemma with $E' = A$ and $E'' = A'$ for each factor

\vspace{.15in}

\noindent $A(1) \ra (\ldots (A(t) \ra p) \ldots )$ of $A$

\vspace{.15in}

\noindent there exists a factor

\vspace{.15in}

\noindent $B'(1) \ra (\ldots (B'(t) \ra p) \ldots )$ of $B'$

\vspace{.15in}

such that $A(i)$ conv. $B'(i)$ for $i = 1, \ldots , t$.  Again by the complete invariants lemma with $E''  = B'$ and $E' = B$ each factor

\vspace{.15in}

\noindent $B'(1) \ra (\ldots (B'(t) \ra p) \ldots ) $ of $B'$

\vspace{.15in}

\noindent there exists a factor

\vspace{.15in}

\noindent $B(1) \ra (\ldots (B(t) \ra p) \ldots )$ of $B$

\vspace{.15in}

\noindent and expressions $C(1) , \ldots , C(t)$ such that $B(i)$ conv. $B'(i) \wedge C(i)$.
End of proof.
\end{proof}

We conclude that $B \subseteq A \Lra$ for each factor $A(1)\ra (\ldots (A(t) \ra p) \ldots )$ of $A$ there exists a factor $B(1) \ra (\ldots(B(t) \ra p)\ldots )$ of $B$ such that $A(i) \subseteq B(i)$ for $i=1,\ldots , t$.  Now we present a polynomial time algorithm for determining whether $A$ conv. $B$. A different algorithm is proposed in \cite{5}.  Clearly, it suffices, given an expression $A$, to determine in polynomial time whether any two subexpressions are interconvertible.  We suppose that the binary tree $A$ has $n$ nodes and these are numbered by depth first search so subexpressions of $A$ have lower numbers than their subexpressions. We construct an $n \times n$ Boolean matrix whose $(i,j)$ entry is $1$ if the $i$th node of $A\ \subseteq $ the $j$th node of $A$ and is $0$ otherwise. We shall fill in the $n \times n$ entries in time polynomial in $n$.  We suppose that we wish to fill in the entry $(i,j)$ and that the entries filled in  for all pairs $(k,l)$ with $k+l > i+j$.  Let $B$ be the $i$th subexpression of $A$ and $C$ the $j$th.  The factors of $B$ are in 1-1 correspondence with its strictly positive atoms;  similarly for $C$.  For each pair of factors $B(1) \ra ( \ldots (B(t) \ra p)\ldots )$, $C(1) \ra ( \ldots (C(t) \ra p) \ldots )$, we consider the entries for pairs of nodes corresponding to the pairs of expressions $C(k),B(k)$, for $k = 1, \ldots ,t$, already in the matrix.  This takes time $O(n^3)$.  If for each factor of $B$ the procedure succeeds for some factor of $C$ we enter a $1$ in $(i,j)$; otherwise, we enter a $0$.  The entire algorithm runs in time $O(n^5)$.  It is correct by beta soundness.
\end{enumerate}

\end{document}